\documentclass[12pt]{article}
\usepackage{amssymb}
\usepackage{amsmath}
\usepackage{latexsym}
\newcommand{\fig}[1]{\includegraphics[width=12cm]{#1.eps}}
\usepackage{graphicx}

\begin{document}

\title{Black hole entropy: inside or out?}
\author{Ted Jacobson, Donald Marolf,  and Carlo Rovelli\\
\noalign{\vspace{3ex}}\\
}

\date{}
\maketitle

\begin{abstract}
A trialogue.  Ted, Don, and Carlo consider the nature of black
hole entropy.  Ted and Carlo support the idea that this entropy
measures in some sense ``the number of black hole microstates that
can communicate with the outside world.''  Don is critical of this
approach, and discussion ensues, focusing on the question of
whether the first law of black hole thermodynamics
can be understood from a statistical mechanics point of view.
\end{abstract}

\newpage

\noindent {\large \bf A Trialogue}

\bigskip

\noindent The following is our rough reconstruction of a discussion
that took place during the summer of 2004.  The participants are Ted
Jacobson, Don Marolf, and Carlo Rovelli.  The topic is the
interpretation of the Bekenstein-Hawking entropy $S_{BH}$ of a black
hole.  The conversation takes place in three scenes, during the course
of an afternoon and an evening.

Ted and Carlo support the idea that the Bekenstein-Hawking entropy
measures in some sense ``the number of black hole microstates that
can communicate with the outside world.''  Don has certain
sympathies in this direction, but does not see a complete picture
in which he can believe.  As a result, he finds the contrasting
point of view compelling.  The contrasting point of view is that
$S_{BH}$ counts the total number of black hole microstates,
including all configurations of the interior.  As the conversation
progresses, all sides attempt to learn something new about this
old, but still exciting, question of black hole
physics.\footnote{In the spirit of an informal discussion,
citations are used only to help the reader locate arguments
specifically referenced in the discussion.  However, all
participants implicitly draw heavily on the work of many other
physicists. Complete references can be found in those few works
cited below.}

\bigskip

\begin{center}
------------------------------------------------------------
\end{center}

\bigskip
\begin{center}
\noindent {\bf Scene I: Entanglement, outside on the terrace}
\end{center}

\bigskip

{\it The setting:  A fine summer afternoon.  A group of physicists
stand on a terrace overlooking a restored 13th century village in
the French Alpes-de-Haute-Provence. }

\bigskip
\begin{center}
------------------------------------------------------------
\end{center}
\bigskip

{\bf Don:}   Greetings, Ted!  It's a pleasure to see you, as
always.  But this time particularly so:  I have a question that I
have been saving for you, and also for Carlo, when I can catch
him! I can see that this is the perfect setting to spring it upon
you.

To begin, I know that you advocate the view that black hole
entropy has nothing to do with the internal states of a black
hole, but is rather a measure of the number of states, associated
with the existence of the horizon, which can influence the outside
world.   That is,  if one could count the complete set of
microstates associated with a given black hole (and including a
description of the interior), you both believe that the correct
result would be far larger than the number $e^{S_{BH}}$ naturally
associated with the Bekenstein-Hawking entropy.

The funny thing about your point of view is that it would remove black
holes from the context of familiar statistical mechanics.  This means
that a number of the usual stat mech arguments would not apply
directly to black holes.  Yet, the original motivation for the
Bekenstein-Hawking entropy is that it played {\it precisely} the role
one would expect for the entropy of a standard statistical mechanical
system.  So, what I want to know is how, within your point of view,
could one possibly explain the laws of black hole thermodynamics?
Here I am in fact most concerned about the {\it first} law of black
hole thermodynamics, which states that two nearby equilibrium
configurations are related by
\begin{equation}
\label{Law1}
dE = T dS + {\rm Work \ Terms}.
\end{equation}
However, for completeness I'll ask you to explain the second law as
well:
\begin{equation}
\label{Law2}
\frac{dS}{dt} \ge 0.
\end{equation}

\bigskip

{\bf Ted:} I don't know why you say my point of view removes black
holes from the context of statistical mechanics. The outside
observers do their stat.\ mech.\ calculations with all the
information available to them. The states strictly behind the
horizon are irrelevant for that. How this different from any other
application of statistical thermodynamics?

\bigskip

{\bf Don:} Well, in the {\it usual} setting, the first law is tied
to the fact that if an energy $dE$ is added to a system in
equilibrium, and the system returns to equilibrium, the energy is
distributed over all the available states. The entropy change dS
reflects the change in the number of microstates after this
re-equilibration occurs.

\bigskip

{\bf Ted:} Exactly! This supports the argument that the states
behind the horizon are irrelevant: the spacetime behind the
horizon is not static, but rather quite time-dependent. Even for a
static black hole, the Killing field is spacelike, not timelike in
there. Everything falls towards the interior, into the
singularity. There seems to be no reason to imagine that any kind
of equilibration process occurs.\footnote{This point was
emphasized in Rafael Sorkin's article \cite{RS1} in the Chandra
volume.}

\bigskip

{\bf Don:} Exactly! I'll concede the point that the interior of a
black hole does not appear to reach equilibrium in any sense we
understand.  But if we take this at face value, then the black
hole is an {\it essentially} non-equilibrium system!  So, why
should the first law hold at all?

\bigskip

{\bf Ted}:
I think equilibrium is a question of timescales.
The black hole horizon appears to
static observers like a thermal state, just as in flat spacetime the
Minkowski vacuum appears as a thermal state to uniformly accelerated
observers.
Because of the gravitational redshift, the temperature of that state
goes to
infinity at the horizon. An infinite temperature corresponds to
infinite energies,
hence infinitesimal times. Thus the equilibration time for the energy
$dE$
crossing the horizon is, in a sense, zero. And voil\`a, the first law
emerges!

\bigskip

{\bf Don:}  Woah. Slow down a minute here.... I didn't catch that
at all. First, when you say that the ``equilibration time \ldots
is \ldots zero," you mean that equilibrium is indeed reached, and
that in fact it happens infinitely quickly?

\bigskip

{\bf Ted:} Yes. Well, I think this is only a very good
approximation, to leading order in the ratio of the Planck length
to the size of the black hole. I imagine that black hole entropy is finite
because some kind of Planck scale cutoff exists on the field
degrees of freedom, and because the Planck length is so short
compared to the other relevant scales, the entropy is dominated by
Planck scale degrees of freedom, an hence it is only their
equilibration that really matters. And their equilibration takes
place over Planck times, in some sense.

\bigskip

{\bf Don:}  Hmmm....  I'm not sure yet how this question of
timescales is relevant.  So, let me put that aside for a moment. I
will certainly agree that, at least in a certain sense, the
exterior of a black hole reaches equilibrium.

However, a greater concern is that the exterior of the black hole
does not exist in isolation. Instead, it is coupled to the
interior of black hole. And we have agreed that the interior does
not reach equilibrium, so why should we be able to apply the rules
of equilibrium thermodynamics to the exterior? In other words,
just {\it how} can we conclude that the first law will emerge?

\bigskip

{\bf Ted:} Well {\it that's} a fat
pitch\footnote{http://www.enlexica.com/sp/bb/index.html}!
As our numerical relativity colleagues gratefully point out,
the exterior evolution is independent of what happens inside
the black hole. The ``coupling" between the interior and the
exterior happens, in effect, just at the horizon, again at
this very short length scale where the only thing happening
is the vacuum fluctuations.

\bigskip

{\bf Don:} I don't think it's so trivial.
After all, isn't it true that the {\it motivation} for your
belief that $S_{BH}$ does not count the total number of
black hole states is that you would like the black hole to
be able to absorb an infinite amount of information, so that
in turn information can be ``lost" inside a black hole as it
evaporates?

\bigskip

{\bf Ted:}  Well, no, that's not my {\it motivation}, but I do
believe the
black hole can absorb lots of information.

\bigskip

{\bf Don:}  Then I think you're trying to have your cake,
and eat it too.  Such information transfer to the interior
requires that we speak about the black hole horizon as being
coupled to the interior.

\bigskip

{\bf Ted:}  Ah, I thought you were referring to coupling the other
way; you're right of course that the exterior can affect the
interior\ldots

\bigskip

{\bf Don:} Let me draw a Carter-Penrose diagram or two to
illustrate the point. ({\it Don takes out a pen and paper and
starts to draw Figure 1, below.})  Let's consider a spacetime
which has a future horizon $H$.   I'll draw this as a heavy line.
Hmm..., I guess I'll draw the familiar asymptotically flat case,
though I don't think this will be important to the story. Let me
also assume that we constantly send a small stream of energy into
the hole to balance the outgoing Hawking flux, so that the black
hole does not evaporate.

Now, there are basically two senses in which we could attempt to
discuss the physics near the black hole... where by ``physics'' I
mean  both the issue of information loss and the approach of a
black hole to equilibrium.  These correspond to looking at two
different families of spacelike hypersurfaces in this spacetime.
This will be clearest if I make two copies of my diagram, with
each copy showing one of the two families of hypersurfaces. I'll
draw the hypersurfaces with dotted lines.

\begin{figure}\label{slices}
\begin{center}
\fig{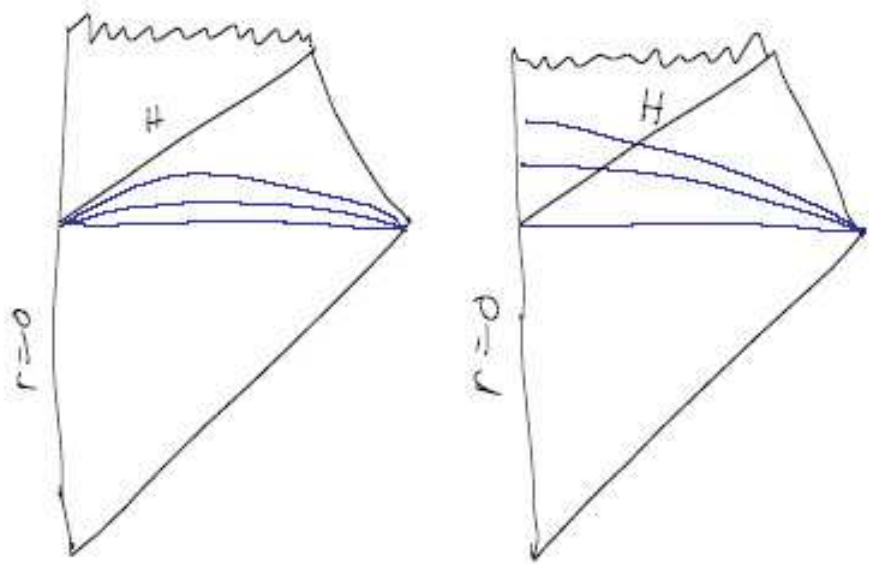}
\end{center}
\caption{Don draws two Carter-Penrose diagrams, each showing a
family of Cauchy surfaces.}
\end{figure}

 On the left, each hypersurface is a Cauchy surface, and each
lies completely in the exterior of the black hole.   I certainly
agree that, so far as we know, physics in this exterior region can
be described as uncoupled from the black hole interior. However,
there are two reasons why I don't think this is the picture you
wish to discuss. First, it is not clear in what sense the black
hole ``approaches equilibrium" with respect to the time evolution
described by this family. Indeed, any sort of junk which falls
into the black hole is present on every surface in this family and
does not disappear. Instead, it moves ever closer (and ever more
slowly) toward the black hole.  Second, this family of
hypersurfaces clearly loses no information.

In contrast, the family on the right is drawn so that each
succeeding hypersurface intersects the horizon a bit farther up.
I've chosen these to be Cauchy surfaces as well, though clearly
information is ``lost" to the interior of the black hole. In terms
of this family, I do see that the black hole approaches
equilibrium after being perturbed: this is just the statement that
the spacetime approaches some stationary solution as one moves up
along the horizon.

 However, we see
that evolution in the sense described by these surfaces includes
both the interior and the exterior regions. In particular, we
should note that there is an exchange of energy across the
horizon: positive energy falls across the horizon passing through
any surface system and into the interior and, due to Hawking
radiation, so does negative energy! Thus, energy is shuffled back
and forth.

So, now, let me restate my original question: Classical general
relativity tells us that if we actively probe a black hole, then
the combined interior-surface system responds in accord with the
first law of thermodynamics.
$$dE = TdS + {\rm Work \ Terms}.$$
But how can this result be derived from your picture in which $S$
is the entropy of only the surface system, and thus does not
include the interior?

\bigskip

{\bf Ted:} The interior is causally disconnected, hence irrelevant
for the outside observer. I would assert that the entropy in the
first law is the log of the number of states of the black hole
that can affect the exterior.

I believe this answers the part of your question regarding how the
interior could possibly be left out of consideration in deriving the
first law,
but it does not yet address how one could {\it derive} the first law, 
whether it be with or without the interior.
I suspect you may be willing to accept my rationale for the
irrelevance
of the interior if I can actually account for the emergence
of the first law without the interior playing any role.

\bigskip

{\bf Don:}  Now, I wouldn't want to overly commit myself in
advance... But if you can explain the first law in terms of the
entropy of your ``surface system" alone, then my question will
indeed have been answered.

\bigskip
{\bf Ted:}  Well, the usual derivations of the first law, 
whether by purely classical {\it mechanical} calculations, or by formal
semiclassical quantum gravity calculations of the thermal partition 
function, do not refer to the
interior. But I suppose you are asking for a ``statistical
thermodynamics" type derivation, where the energy falling across
the horizon is directly linked to the change of the number of
microstates, right? ({\it Don nods, smiling and raising his
eyebrows.}) Hmm, lemme see here\ldots

Well, then, I'm going to have to commit myself to a candidate for
the microscopic states. I'll profess to the belief that these are
the vacuum fluctuations just outside the horizon, which carry
entropy since they are correlated to fluctuations inside the
horizon. That is, I think the black hole entropy is vacuum
entanglement entropy\footnote{Ted recommends the reference
\cite{RS1}.}. The fact that this scales with the horizon area
follows simply from dimensional considerations. Thus there is a
universal proportionality constant relating horizon area to
entropy. The classical proof of the first law, together with the
Hawking temperature $\hbar\kappa/2\pi$, then implies the identity
of this proportionality constant as being precisely $1/4\ell_{\rm
Planck}^2$.

\bigskip

{\bf Don:}  I'm not at all sure there are enough states in such
fluctuations to do the job\footnote{Don refers to arguments in
\cite{HF}.}, but that's probably a side issue for our present
discussion.

\bigskip

{\bf Ted:} Hey, to me that sounds crucial for the main issue, not
a side issue at all! If you're right that there are not enough
states outside the horizon, then there's no way my picture could
be correct.  We should discuss this too!

\bigskip

{\bf Don:}  That would be great.... I would love to have a
separate discussion with you sometime about all of my concerns
regarding the ``entanglement entropy scenario," but I think we'll
have plenty to discuss today just sticking to broader questions
about the first law....

Here, the main point for me is that you haven't yet gone far
enough. What I am asking is how, microscopically, is the energy
thermalized into your horizon degrees of freedom, in accordance
with the first law.

\bigskip

{\bf Ted:} Maybe we are making progress here, since the question
has now focused down onto the equilibration process\ldots Well, if
the entropy arises from the thermal nature of the vacuum outside
the horizon as viewed by uniformly accelerated observers, all I
really need to explain is why the vacuum remains the
vacuum at
the short distance scales where the states counted by the
Bekenstein-Hawking entropy reside. My answer would be that,
because of decoupling, the vacuum is not excited at these short
scales.

\bigskip

{\bf Don:} I don't see how to get $dE=TdS$ from this statement.
In particular, when the energy $dE$ thermalizes it must cause {\it
some} change, or else the entropy would not grow!

\bigskip

{\bf Ted:}  I guess my argument has been somewhat cryptic so far.
Let me try to remedy that. To begin with, I think it would help to
be very concrete about what you mean by deriving the first law
from a microscopic description. In a typical setting of a system
in thermodynamic equilibrium, the statistical description of the
system is given by a density matrix $\rho$. Let's fix on the
canonical ensemble, so $\rho=\exp(-\beta H)/Z$, where $\beta$ is
the inverse temperature, $H$ is the Hamiltonian, and $Z$ is the
trace of $\exp(-\beta H)$. Suppose some energy is added to the
system, but in such a way that $\delta\rho\ll\rho$.  Then, as
discussed in your paper\footnote{Ted refers to \cite{Notes}.} with
Minic and Ross,
 varying the entropy $S=-{\rm Tr}(\rho\ln\rho)$ and the
mean energy $\langle E\rangle={\rm Tr}(\rho H)$, one
finds the relation
\begin{equation}
\delta S = \beta\, \delta\langle E\rangle,
\end{equation}
hence the thermodynamic relation $dE = TdS$ that we have been
calling the first law.\footnote{Actually, the ``first law" should,
I think, be the statement of energy conservation in thermodynamics.
In this relation $dE$ refers to the
heat $\delta Q$ added to the system, and I don't know a name for
the relation $\delta Q = TdS$! Nevertheless, people seem always to
call this the first law in the context of black hole
thermodynamics, so let's keep doing so.} Thus once we have a
system whose density matrix is a canonical ensemble, we have a
derivation of the first law, agreed?

\bigskip

{\bf Don:}  Yes, Ted, I agree\ldots provided that you can justify
the statement that your system is in a canonical ensemble,
especially {\it after} you make the change $\delta \rho$\ldots and
provided that we are discussing the physically relevant system.

\bigskip

{\bf Ted:} Well then here's my argument. First,  I claim we know
that the ``surface system" is well described by a canonical
ensemble, as long as it is very near the local vacuum state.
Here's the reason: at short distances a little patch of the black
hole horizon may be identified with a patch of a Rindler horizon
in Minkowski spacetime. The Unruh effect, or what is known in
algebraic quantum field theory as the Bisognano-Wichmann theorem,
tells us that, as viewed from one side of a Rindler horizon, the
Minkowski vacuum is a canonical ensemble at dimensionless
temperature $T=1/2\pi$ with respect to the dimensionless boost
Hamiltonian (i.e. the generator of Lorentz boosts). This theorem
holds for any interacting field theory on a Minkowski background.
The proofs of this fact are very deep, relying only on stability
of the vacuum and local Lorentz symmetry, so I would assert that
there is good reason to think that it extends to quantum gravity.
The system of quantum field fluctuations at short distances
outside a local Rindler horizon is well described by a canonical
ensemble, and it is the collection of these systems outside a
black hole horizon that I propose to identify with the ``surface
system" of a black hole. The entropy of this ensemble is the
entanglement entropy of the the vacuum, and is divergent in
ordinary local quantum field theory due to the absence of a UV
cutoff. I suppose that it is somehow rendered finite in quantum
gravity.

Now suppose some energy crosses the black hole horizon.
 It is the thermodynamic description
used by the outside observers that interests us. These observers
never see the energy cross the horizon, they just see it approach
ever closer. That is, they see it sink into the surface system.
Since the surface system is initially described by a canonical
ensemble, and since the energy crossing the horizon is assumed to
make only a very small perturbation of the vacuum at short
distances, the absorption of this energy into the surface system
is associated with an increase of entropy $\delta S = \delta E/T$.
Here $E$ is the mean value of the boost Hamiltonian,  and $T$ is
the dimensionless temperature $1/2\pi$. As pointed out in your
paper\footnote{Ted again refers to \cite{Notes}.}, the variation
of the entropy is finite even though the entropy itself is UV
divergent. One can impose a UV cutoff, compute the variation, and
take the cutoff away, which yields a variation independent of the
cutoff. Even with a cutoff in place, one would obtain the same
result, provided the cutoff is at a large enough energy.

So far I've just used the local Rindler horizon approximation to a
patch of a black hole horizon.  Now I put together such patches to
cover the black hole horizon, rescaling the boost Killing vector in
each patch to agree with the black hole horizon generating Killing
vector.  This also rescales the boost Hamiltonian and temperature by
the surface gravity of the black hole so that they agree with the
Killing energy and Hawking temperature.  (In the case of a rotating
black hole it is actually the Killing energy minus the Killing angular
momentum times the angular velocity of the horizon.)  This yields the
relation $\delta S = \delta E_{BH}/T_H$, where $E_{BH}$ is the black
hole energy and $T_H$ is the Hawking temperature.  The argument up to
here is, I think, essentially equivalent to the one spelled out in
detail by Zurek and Thorne using the hypersurfaces on your left hand
diagram.\footnote{Ted refers to \cite{MP}.} This is not yet a
derivation of the first law for black holes, since in that law what
appears is the variation of Bekenstein-Hawking entropy
$S_{BH}=A/4\ell_{\rm Planck}^2$, not the variation of entanglement
entropy.

But we are almost there.  As I said before, the entanglement entropy
is proportional to the surface area, with a proportionality constant
$\eta$ that seems quite plausibly universal, since all local Rindler
horizons are equivalent.  Thus it only remains to explain why
$\eta=1/4\ell_{\rm Planck}^2$.  One might just refer to the classical
geometric derivation of the first law in general relativity, and
``read off" this relation by comparison with the statistical first
law, as did Zurek and Thorne.  But I think this would not be
fulfilling your call for a truly statistical derivation of the first
law for black holes.  To establish this link between $\eta$ and the
Planck length, I will invoke my derivation of the Einstein equation as
a thermodynamic equation of state\footnote{Ted refers to \cite{TOS}.}
.  In that paper, I showed that the validity of the first law for an
entropy of the form $S=\eta A$ at all local Rindler horizons implies
that the Einstein equation {\it must} hold, with Newton's constant, or
rather $\hbar G = \ell_{\rm Planck}^2$, equal to $1/4\eta$.

I believe that this line of reasoning not only provides a microscopic
derivation of the first law of black hole thermodynamics, but it makes
abundantly evident the fact that the interior of the black hole is
irrelevant
to the story.

\bigskip

{\bf Don:}   Thanks, Ted. I think we are now moving to the heart
of the matter.... For me, the most important thing you just said
was

\begin{quote}
``It is the thermodynamic description used by the outside
observers that interests us. These observers never see the energy
cross the horizon, they just see it approach ever closer. That is,
they see it sink into the surface system.''
\end{quote}

So then, it seems that we have now changed the framework of the
discussion to one associated with the {\it left} diagram that I
drew. ({\it Don points again at Figure 1, and taps the left-hand
diagram demonstratively.}) In this setting I could indeed imagine
a derivation of the first law.  But now what does it mean to say
that your surface system does {\it not} describe the entire black
hole, but that in addition there is a separate interior system?

In particular, doesn't the data near the horizon determine the
interior of the black hole?  And doesn't this mean that your
surface system must have at least as many states as the black hole
interior?

\bigskip

{\bf Ted:} Hmmm\ldots mmm\ldots hold on, no. The statement that
outside observers never see the energy cross the horizon does not
entail any choice of surfaces. And I really don't want to use the
surfaces that never cross the horizon, since on them there is no
entanglement entropy.

Let's call the surfaces on the left
``static surfaces", and those on the right
``horizon crossing surfaces". It is true that a packet
of  infalling energy does disappear from the exterior portion
of the horizon crossing surfaces after some time.
However, this presents no problem for my derivation
of the first law, so what are you concerned about?

\bigskip

{\bf Don:}  Perhaps my asking about hypersurfaces was not helpful,
but let me try to phrase the question another way. Consider a
process in which an energy $dE$ is added to the black hole.  It is
reasonable that this energy first enters your surface system, but
the question is then about what happens next: Does the energy $dE$
remain within the surface system and thermalize there? Or does
part (or all) of the energy enter the interior?

Now, back when you referred to a description used by outside
observers, I thought that you wished to say that the energy $dE$
remained in the surface system.  In this context, I find it
plausible that the energy might thermalize and lead to $dE= T
dS_{surface}$. Mind you, this thermalization does not appear to
occur classically, but I am happy to interpret the results of, for
example, my paper with Minic and Ross that you mentioned earlier
as evidence that such a thermalization does occur quantum
mechanically.

However, if I use a description where all of the energy added to
the black hole stays within the surface system, then I don't see a
sense in which I can talk about the black hole interior as a
separate system carrying additional degrees of freedom.  Can you
fill me in?

\bigskip

{\bf Ted:}  What seems to bother you is that you think I claim
that on the one hand the energy $dE$ thermalizes in the surface
system, but on the other hand it crosses the horizon. This sounds
impossible, since to say that energy thermalizes is to say it is
distributed over a large number of a specified set of degrees of
freedom, which locates the energy in those degrees of freedom, and
you think I am saying that the energy is both in the surface
system degrees of freedom and in the interior degrees of freedom
at the same time (i.e. on the same horizon crossing slice), which
is a contradiction since the energy cannot be in two places at
once. Is that an accurate description of your objection?

\bigskip

{\bf Don:} Yep.

\bigskip

{\bf Ted:} Well if you look back you'll see that I never said the
energy thermalizes in
the surface system! What I said, just after what you previously
quoted, was
\begin{quote}
``Since the surface system is initially described by a canonical
ensemble,
and since the energy crossing the horizon is assumed to
make only a very small perturbation of the vacuum at short distances,
the absorption of this energy into the surface system is associated
with
an increase of entropy $\delta S = \delta E/T$. Here $E$ is the mean
value of the boost
Hamiltonian,  and $T$ is the dimensionless
temperature $1/2\pi$."
\end{quote}
The subtlety here is that the energy may be macroscopic, e.g. a
pebble, while the
entropy change $\delta S$ is carried by the near Planck scale degrees
of freedom.
I am not asserting that the pebble mass is spread over these degrees
of freedom.
So ``thermalization" of the pebble in the sense you were concerned
about does not,
and need not, occur in order for me to invoke the formula
$\delta S =\delta E/T$.

Now you might object that, in the description I wish to use,
the pebble energy does not remain in the surface system,
but rather crosses the horizon into the black hole. And this
suggests that the surface system has then {\it lost} the pebble
energy. So, you might well ask, why does the surface system
entropy not go back down to where it started before the pebble
arrived?

I claim that the answer lies in gravity. In a non-gravitational,
flat spacetime setting, the pebble would pass through the Rindler
horizon surface system  and out the other side, and the entropy
would go back down. But you {\it cannot} turn off gravity. The
process of the pebble falling into the black hole is associated
with an increase in the mass of the black hole, which therefore
acquires a slightly larger horizon area.

In a bit more detail, the relation between the pebble energy and
the surface system equilibrium is as follows. The pebble is much
wider in the radial direction than the surface system, so only a
very small part of the pebble energy resides in the surface system
at any given time. The passage of pebble energy makes a small,
adiabatic perturbation on the surface system, increasing the
entropy. When that energy exits the surface system and crosses the
horizon, it leaves behind an entropy imprint via the increase of
horizon area. The surface system is then in a new canonical
ensemble, since it is a slice of the near-horizon vacuum on the
background geometry of a slightly larger black hole. Hence the
increase of surface system entropy is for keeps (unless we let the
black hole evaporate), even though the pebble energy falls across
the horizon and does not remain in the surface system.

\bigskip

{\bf Don:}
 Very good, Ted. I now see your picture much more
clearly. But do you think that your arguments above really
constitute a {\it derivation} of the first law from a complete
statistical mechanical picture? Or, at this stage are you instead
{\it using} the classical first law to deduce that ``somehow"
gravity leads to the scenario for ``thermalization" described
above in which the surface system gets the entropy ``for keeps"?

\bigskip

{\bf Ted:} I think the argument really constitutes a derivation
of the first law from a statistical mechanical picture. The fact
that classical gravity enters in a critical way to ``preserve" the
entropy change in the surface system is just a peculiar
feature of this physical system, but does not make the
derivation any less statistical mechanical.

\bigskip

{\bf Don:}  Well didn't you appeal earlier to your derivation of
the Einstein equation {\it from} the classical first law in order
to relate the entropy change of the surface system to the change
of the Bekenstein-Hawking entropy of the black hole?

\bigskip

{\bf Ted:} Remember, the statistical argument gave me a first law
involving the entropy change $\eta\, \delta A$. I then argued that
this is consistent
for all local Rindler horizons only if there is gravity and it acts
according to the Einstein equation with Newton's constant equal
to $1/4\hbar\eta$. So I don't think I've assumed that the first law
holds.
I've argued statistically that it holds with entropy $\eta A$,  and
inferred
from that the existence of gravity with the above Newton constant.
Thus I {\it deduce} that this is in fact the Bekenstein-Hawking
entropy.

\bigskip

{\bf Don:}  OK, let me see if I've got the argument now.  You
first {\it assume} that your model is consistent, and in
particular that there must be some mechanism to (permanently)
increase the entropy (and thus the area) of your local Rindler
horizon.  Given this assumption, it then follows that this effect
must be the same as one would derive from the gravity and the
Raychaudhuri equation.  In fact, you actually get all of the
Einstein equations from this argument!

\bigskip

{\bf Ted:} I hadn't realized so, but yes, I have assumed that the
entropy increase is permanent (neglecting Hawking radiation). 
The area increase on the other hand is not an assumption but is inferred.

Perhaps the argument can be improved, however. It seems
the permanence of the entropy increase of  the local Rindler horizon
is {\it required} if  the second
law of thermodynamics is to hold for the outside observers!
Such a version of the second law would refer to the total entropy of the
exterior, including the surface system.
Let's call this the ``exterior second law", or ``ESL" for short.

\bigskip

{\bf Don:}     Aha!  So, you wish to kill two birds with one
stone! Well, one concern I have here is that, at least in standard
statistical mechanics, the first and second laws have a very
different microscopic character.  The second law of course holds
only in a statistical sense, whereas the  first law is a precise
statement about ensembles.  That is, if a warm lump of coal
happens to undergo a fluctuation to a lower entropy state (in a
microscopic violation of the second law), then this is certainly
accompanied by the emission of a small bit of energy of magnitude
roughly $T dS$ (and thus satisfying the first law)..... But this
is a fine enough point that I suppose you can claim that both
features will be enforced by the same microscopic mechanism.

On the other hand, I find it hard to see how {\it any} microscopic
mechanism would enforce the second law in your setting... I mean,
we don't usually say that the region {\it outside} of a black hole
satisfies the second law at all. Rather, we say that we have a
{\it generalized} second law, where entropy falling across the
horizon leads to an increase in black hole entropy.  I realize
that in your scenario you want the generalized second law to
become the regular second law for your surface system together
with the exterior.... But what I want to emphasize is that you
need the surface plus exterior system to satisfy the second law in
the form $\delta S \ge 0$ usually reserved for {\it closed}
systems. And you need this despite agreeing that energy can pass
into the interior, so that this system looks to me to be an {\it
open} system, which would naturally satisfy a second law of the
form $\delta S \ge Q/T$, which allows $S$ to decrease when $Q$ is
negative; i.e., when heat leaves the system. I emphasize that this
weaker form of the second law would not be sufficient for your
purposes.

\bigskip

{\bf Ted:} It's interesting how we have arrived at the very issue
that led to the birth of black hole thermodynamics. I think it was
precisely the concern that the second law would be violated for
outside observers that led Bekenstein to propose that black holes
have entropy! While the exterior is an open system in the sense
that things can leave it, it is self-contained dynamically
(since the horizon is a causal barrier), which
is not true for a generic open system.  If the ESL did {\it not} hold,
I'd expect that one could make a
perpetual motion machine exploiting that.

It has often been remarked that a failure of
the {\it generalized} second law---for black holes
or even for de Sitter spacetime---would lead to
the possibility of  constructing a perpetuum mobile. Such
remarks assume that the Bekenstein-Hawking entropy $S_{BH}$ is
the balance of the ``real" entropy in the system, so that standard
thermodynamic reasoning can be applied.
I am saying the same thing, but in place of
$S_{BH}$ I have the surface contribution to the entropy visible to
the outside observers. Since the surface
contribution {\it is} the balance of real entropy in the system
accessible to the outside observers,
the perpetuum mobile argument should be
valid for the ESL.

\bigskip

{\bf Don:}  I'm quite skeptical
\ldots So, I'd really like to see you try to write
down the details.... In any case, I think
you are implicitly making another assumption, namely, that the
horizon is a strict causal barrier even at the quantum level, so
that no entropy from inside can ever influence the exterior.

\bigskip

{\bf Ted:} Yes. it is an underlying assumption of my picture that
no information from inside can influence the exterior. In this
setting, I think the perpetuum mobile argument strongly supports
the ESL, although it is not the {\it microscopic} account you are
asking for. Rafael Sorkin has thought a lot about precisely this
question, and has offered an outline of a microscopic proof in
both the semi-classical and full quantum gravity
settings.\footnote{The ideas are explained in Refs.
\cite{RS1,RSGSL}.} I think there are gaps in the outline, but
maybe he's on the right track.

\bigskip

{\bf Don:} Maybe, but maybe not \ldots

Well, Ted, even after all of this discussion
I'm afraid I still find it very hard to see how any microscopic
mechanism could allow your surface system to ``keep'' the entropy
associated with the passage of a bit of energy $dE$ after this
energy has passed on through the horizon. I guess we'll have to
leave the issue issue here for now, but I would have to see this
sorted out in detail before I could really accept your point of
view.

\bigskip

{\bf Ted:}  I'm not sure if {\it I} accept my point of view. Your
questioning has forced me to sharpen my views about the nature of
black hole entropy into a more definite picture. For the moment,
it makes complete sense to me, and in particular I think it gives
a cogent answer to why the surface system keeps the entropy. But
our discussion has certainly given me much food for thought.

\bigskip

{\bf Don:} Well, then, I am sure we will discuss this again! But,
speaking of food, it looks like it's now time for dinner. Shall we
adjourn to the dining room? Maybe we can get Carlo to join
us\ldots

\bigskip
\begin{center}
------------------------------------------------------------
\end{center}
\bigskip

\begin{center}
\noindent {\bf  Scene II:  Horizon fluctuations over dinner}
\end{center}

\bigskip
 {\it Don and Ted join Carlo at the dinner table.  The aromas
of a Provencal meal in preparation waft in from the kitchen.}

\bigskip
\begin{center}
------------------------------------------------------------
\end{center}
\bigskip

\bigskip

{\bf Don:}  Hello, Carlo!  I've been grilling Ted over his views
on black holes and the idea that the Bekenstein-Hawking entropy
resides in a set of surface states.  In particular, I've been
asking how he would derive the first law in this picture.  You are
also known to expound this view, so I'd love to get your answer as
well!

\bigskip

{\bf Carlo:} Indeed, I was hoping to catch you!  I was listening to
your conversation earlier on the terrace, but I did not want to break
your line of discussion.  Now it seems that I have my chance!

My general understanding of the situation is very much in line with
Ted's.  I agree with Ted that we can understand the BH entropy in
terms of degrees of freedom sitting on the surface.  The pebble falls
across the horizon and leaves an ensemble with higher entropy.  I also
agree with Ted that we can conceptually identify three subsystems
here: the exterior, the surface and the interior, and that the key of
the story is in the peculiar property of the interaction between the
surface and the interior: the evolution of the interior depends on the
surface, but not viceversa: the evolution of the state of the surface
(and the exterior) does not depend on the state of the interior.

\bigskip

{\bf Ted:} So far, quite good\ldots

\bigskip

{\bf Carlo:} But there are also differences in the way you and I look
at the problem, Ted \ldots

\bigskip

{\bf Ted:} No doubt!

\bigskip

{\bf Carlo:} Your picture is in terms of fields on a spacetime
background.  This spacetime background is taken to be non-fluctuating
neither quantum mechanically nor thermally.  Such a picture is
necessarily approximate.  I want to consider a picture which is
physically more complete, in which we take into account that the
spacetime geometry fluctuates.  I think that this can provide a key
ingredient to better understand the entropy of the hole.

The spacetime geometry fluctuates for two reasons.  One is quantum
mechanics, the other is just because at finite temperature \emph{all}
dynamical quantities undergo statistical fluctuations.  The two kind
of fluctuations might perhaps even be related.  But
even disregarding quantum fluctuations, spacetime geometry must
fluctuate thermally in a hot environment.  Therefore it is not
realistic to assume a given fixed background geometry, such as, say, a
Schwarzshild background.

\bigskip

{\bf Ted:} I don't think we need to address these questions for the
understanding of black hole thermodynamics, since the setting is that
of a black hole very large compared to the Planck length.  We need
only that the leading order term in the entanglement entropy scales
with the area, with a universal coefficient.  All other details are
strictly beyond the purview of our current understanding of quantum
gravity, and beyond the theoretical resolution needed to understand
the origin of black hole thermodynamics.

\bigskip

{\bf Carlo:} Maybe.  But geometry fluctuations {\em are} there.  I
think that, if we want to do statistical mechanics in a context in
which gravity plays a central role, it may not hurt to actually
consider the statistical fluctuations of the gravitational field,
namely of the geometry.  Notice that this can be done irrespective
of whether we think of the metric as a fundamental variable or as a
collective variable.  About quantum gravity, I respect your view that
the quantum gravitational picture of the situation is beyond the
purview of our current understanding of quantum gravity, but I do not
share it!  However, this is not the subject of our discussion!

\bigskip

{\bf Ted:} Okay, let's consider the fluctuations of the geometry.
Then?

\bigskip

{\bf Carlo:} Then, the geometry of the horizon fluctuates.  The
horizon is in a thermal mixture of different positions, or different
shapes.  The horizon is never a perfect sphere, but rather a closed
surface that fluctuates over an ensemble of shapes.  I think that the
degrees of freedom responsible for the entropy must be precisely the
ones characterizing the geometry of the horizon itself, namely the
{\it shape} of the horizon.

\bigskip

{\bf Ted:} The horizon is by definition the causal boundary of the
past of future null infinity.  The fluctuations in that boundary today
are determined by the entire future.  So how can one meaningfully talk
about the shape of the horizon at one time as a dynamical variable
that can carry entropy?

\bigskip

{\bf Carlo:} I think that a clean description of the system is
atemporal: the statistical mechanics of a relativistic gravitational
system is defined classically by an {\em ensemble of solutions of the
field equations}.  If we then choose a foliation in each, and (for the
sake of the analysis) identify equal time surfaces, then we get an
ensemble of ``horizon shapes" for each coordinate time $t$. 

\bigskip

{\bf Don:} This sounds quite different from Ted's picture.  You say
that the degrees of freedom counted by the BH entropy are the ones
characterizing the shape of the horizon.  Ted says it is the
entanglement entropy of quantum fields around the horizon.

\bigskip

{\bf Carlo:} Yes, this is indeed a key difference I wanted to point
out.  The picture that I find more compelling\footnote{Carlo refers to
\cite{carlo}.} is in a background independent context, Ted's is in the
non-fluctuating background approximation.  The two pictures are not
necessarily incompatible.  Suppose you accept this idea that the
degrees of freedom counted by the BH entropy are the ones
characterizing the shape of the horizon.  But you decide to describe
the gravitational field in terms of a background plus a fluctuation.
Then the quantum fluctuations of the shape of the horizon get
reinterpreted simply as the fluctuations of the gravitational field in
the vicinity of the horizon, which is what Ted wants to identify as
the source of the entropy.

\bigskip

{\bf Ted:} At the level of effective field theory, certainly the
gravitational field fluctuations are a part of the overall set of
interacting degrees of freedom.  Also any matter field fluctuation
drives a metric fluctuation, so they are inextricably linked,
especially at the Planck scale where most of the entropy presumably
resides.

\bigskip

{\bf Carlo:} Yes!

\bigskip

{\bf Ted:} So, I agree that the horizon fluctuations should be
included as {\it a} source of black hole entropy, but I don't think
they are {\it the only} source.  I don't see how it could be correct
to attribute {\em all} of the entropy to just the fluctuations of the
horizon geometry.

\bigskip

{\bf Carlo:} Mmmm\ldots my reason to limit my consideration to these
is double: first, a vague intuition that gravity ends up dominating at
very small scale; second, the quantum gravity calculations that appear
to be sufficient to give the correct entropy.  But I am happy to leave
the issue of whether the matter contributions are of the same order
open\ldots

\bigskip

{\bf Don:} Good.  Let's put aside for now the question of the
identification of the surface degrees of freedom---shape of the surface
(and thus purely gravitational) for Carlo, or entanglement entropy (of
all fields) for Ted---and of their relations, and let us get back to my
question about the first law.

\bigskip

{\bf Carlo:} Perfect.  It seems to me that the core of your question,
Don, is the following: if the surface degrees of freedom thermalize,
is equilibrium reached by sharing the energy over the surface degrees
of freedom alone, or rather over the degrees of freedom of surface and
interior?  Is this your question?

\bigskip

{\bf Don:} Yes, indeed.

\bigskip

{\bf Carlo:} Your point, if I understand correctly, is that if the
surface thermalizes together with the interior, then the relevant
number of degrees of freedom is the ones of the coupled
surface+interior system.  But if the surface thermalizes just by
itself, then no information is lost inside.  Right?

\bigskip

{\bf Don:} Yes, and in particular, in the second case I don't
really understand what it means to say that there is a separate
interior system with additional degrees of freedom.

\bigskip

{\bf Carlo:} And you object that, contrary to what Ted and I think,
the ``first" law of thermodynamics $dS=dE/T$ cannot be derived, if $S$
is determined by the number of surface states alone.  Is this is a
fair characterization of the disagreement?

\bigskip

{\bf Don:} Yes, Carlo, that's it exactly.

\bigskip

{\bf Carlo:} Well, I think that, to use your expression, we can have
our cake, and eat it too.  So, let me begin by sketching a picture of
a statistical interpretation of BH entropy.  I may refine it and fill
missing steps in as we go along.

I want to analyze the problem classically, namely disregarding quantum
theory.  I will only bring some quantum effects into the picture later
on, when they are relevant.  This, let me say upfront, is the main
weakness of the picture.  I wish I had a fully quantum picture, but I
do not, for reasons that will be clear in a minute.

So, in classical GR (plus any matter you'd like), let me call $\Gamma$
a global ensemble of states of the system, and $\Gamma_i$, $\Gamma_s$
and $\Gamma_e$ the ensemble of the possible states of the interior,
surface and exterior.  As I said before, $\Gamma$ is an ensemble of
solutions of the classical field equation.  It is an a-temporal
statistical state.

If we fix a one-parameter family of horizon-crossing Cauchy surfaces
(like the ones in Don's picture on the right) labeled by a parameter
$t$, then each (atemporal) state in $\Gamma$ defines the three
evolving states $s_i(t)$, $s_s(t)$ and $s_i(t)$.  Now, the future
evolution of $s_s(t)$ and $s_e(t)$ is independent from $s_i(t)$, but
not the past evolution.  Hence there is information loss (we have the
cake).

Now, the ``equilibrium states" are the ones for which certain
macroscopic parameters do not change with $t$.  Suppose at some $t$ we
have a certain (energy) macroscopic parameter $E_e$ in the exterior,
and a certain (energy-like) macroscopic parameter $E_s$ on the
surface.  I consider microcanonical statistical ensembles defined by
all the microstates in which these macroscopic parameters have a given
value.  I assume an ergodic hypothesis that all processes are allowed,
in the sense that the $t$ evolution drives the evolving states all
over this microcanonical ensemble.  A given value of $E_s$ defines the
surface statistical state.

Next, let the number of surface plus exterior states characterized by
$E_s$ and $E_e$ be $N$.  By ergodicity, $N = N_{e}(E_{e}) \times
N_{s}(E_{s})$, where $N_e(E_{e})$ is the number of states of the
exterior with Energy $E_{e}$ and similarly for $N_s(E_{s})$.  Consider
a different macroscopic equilibrium state obtained by an adiabatic
process that decreases $E_{e}$ by $dE$ and increases $E_{s}$ by a
corresponding amount.  The corresponding number of microstates will be
$N(dE) = N_{e}(E_{e}-dE) \times N_{s}(E_{s}+ dE)$.  If our original
state is the most probable, $N$ must be the maximum of $N(dE)$,
obtained for $dE=0$.  Hence $\frac{d}{dE} log N(E) = 0$.  Namely
$\frac{dS_{e}(E_{e})}{dE_{e}} = \frac{dS_{s}(E_{s})}{dE_{s}} \equiv
\frac{1}{T}$.  That is, an equilibrium situation is characterized by a
macroscopic parameter $T$ which must be the same for the exterior and
the surface and such that $dE = T dS$.  Thus, it seems to me that this
relation can be derived without conflict with the fact that
information is lost to the interior.  We can have the cake, and eat
it\ldots

\bigskip

{\bf Ted:} Well, Carlo, you have just reiterated the standard stat
mech treatment of equilibration between two systems, in this case the
surface and the exterior.

\bigskip

{\bf Carlo:} Essentially, yes.  But carefully choosing a version which
is sufficiently general to be used in a generally covariant
context and adapted to our case\ldots

\bigskip

{\bf Ted:} Okay, but I see several problems with this line of
reasoning, since the assumptions of the standard argument don't seem
to hold here.  First, you said the exterior loses energy $dE$ and the
surface gains $dE$.  This assumes we are in a setting where two
systems are exchanging energy and total energy is conserved.  But you
started out saying that the surface parameter $E_s$ is ``energy-like".
Since there is no conservation law for $E_e+E_s$, your calculation of
the condition for maximizing the number of states seems to me
unjustified.

Second, the first law for black holes relates the {\it energy}
change of the black hole to its Hawking temperature. How are you
characterizing the energy of the black hole?  And how are you
relating the $T$ in your equation to the Hawking temperature?

And third, your argument requires that if I throw energy $dE$ into
a black hole, the states of the exterior with energy $E-dE$ are
equiprobable. But this assumes the exterior is in equilibrium with
this energy. Why should that be so? A pebble falling into a black
hole does not look like an equilibrium state. In my argument, the
equilibrium pertains to the vacuum fluctuations at short
distances. You are eschewing the role of these fluctuations in
your account, but then how do you account for the equilibrium you
are invoking?

\bigskip

{\bf Carlo:} Very good.  This takes us to the center of the discussion
and connects to what the two of you were discussing earlier on the
terrace.  Seems to me that the delicate points in your picture,
vulnerable to Don's criticisms were precisely connected to this issue.
Doesn't the energy $dE$ that enters the horizon immediately cross over
to the interior?  And then how can it contribute to the increase
$dS=dE/T$ of the entropy of the horizon's degrees of freedom?  This
was one of Don's main lines of attack.  Your answer made reference, I
believe correctly, to the specific features of the gravitational
dynamics.  \bigskip

{\bf Ted:} Yes, but as I said before after the relentless
questioning by Don, I don't need to assume that dynamics. I need
assume only the second law of thermodynamics, and I deduce the
gravitational dynamics.

\bigskip

{\bf Carlo:} Good.  The picture I am discussing is different because I
take a different starting point.  Let me first sharpen what I just
said.

You say, very truly, that there is no energy conservation that allows
us to follow the usual statistical mechanical line of thinking.  In
fact, a strict attachment to the nonrelativistic notion of energy is
not useful in the general relativistic context, where energy is a more
delicate notion.  Now, what is the role of energy conservation in
the microcanonical setting?  It is to reduce the region of phase space where
the system is free to wander (ergodicity then tells us the systems are
all over the allowed region).  Now, in general relativity there is no
analogous energy conservation.  Thus, we have to look for a dynamical
input from the Einstein equation that can play the same role that
energy conservation plays in the nonrelativistic context.  We have
precisely what we need: the theorem stating that {\em ``the area of
the horizon does not decrease"}.  This theorem captures the relevant
information from the general relativistic dynamics needed for
understanding the statistical mechanics of the horizon.

Now, if the area cannot decrease, and if we are, as we assumed, in a
statistical equilibrium configuration (where everything that
microscopically changes will later change back), then the only
possibility is that the ensemble fills a region of constant area.  In
other words, a microcanonical ensemble determined by the value of the
macroscopic area parameter is singled out by the Einstein dynamics.
This is what replaces the role of energy conservation in the usual
nonrelativistic microcanonical ensemble.

Notice that I have not proven that the surface degrees of freedom do
in fact thermalize.  But this is {\em always} horrendously difficult
in statistical mechanics!  We can just follow the usual argument that
under some ergodicity hypothesis tells us what are the equilibrium
states, if they exist.  In statistical mechanics, the relation
$dS=dE/T$ is derived as a relation between nearby equilibrium states,
not as a description of a dynamical process of approach to
equilibrium...

What I am saying is that the area is the only quantity that
(classically) cannot decrease, and therefore the ensemble is formed by
all the microstates with a certain area.  The Einstein equation tells
us also that if we throw a certain amount of energy $dE$ across the
horizon, then the area will increase by a certain amount.  Hence a
change in area is connected to a change in external energy.  If we
throw an energy $dE$ inside the hole, then its horizon will fluctuate
over a different ensemble, characterized by a larger area.  There is
no need to think that there is an actual ``Energy" that remains on the
surface degrees of freedom without continuing inside the hole.  The
Einstein dynamics allows us to apply the usual statistical mechanics
logic, with just what is needed: a condition on the set of microstates
in which the system can be.  I think that this is an answer to Don's
question about the fact that energy enters the hole and therefore
cannot thermalize on the surface degrees of freedom...

\bigskip

{\bf Ted:} If I understand correctly, you are saying that you need
not invoke energy conservation, since its only role is
to fix the region of phase space in which the system can wander,
and you propose that, because of its classical non-decreasing
property, the horizon area is suitable for playing that role in
the surface system\ldots

\bigskip

{\bf Carlo:} Yes.

\bigskip

{\bf Ted:} But Carlo, I don't see how the classical area theorem
can consistently motivate a classical fixed area microcanonical
ensemble. The reason is that the fluctuations which you certainly
expect to have in your ensemble are certainly not static, yet it
is only in the static limit that there is no area increase: the
slightest amount of gravitational radiation, for example, produces
a Weyl tensor component that generates shear and thus expansion of
the horizon generators.

I'm not saying that I don't believe the horizon area is
fluctuating with some mean value determined by the macroscopic
parameters---I do. But I don't think this is related to the
classical area theorem, nor do I think you have have justified
your microscopic picture of the horizon as a ``system" in a
certain dynamically selected ensemble.

\bigskip

{\bf Don:} I have a related objection, Carlo. Your statistical
argument considered the surface system together with the exterior,
and in particular it considered the most entropic state of the
joint system formed by taking the surface system and exterior
together. What I do not see is how you justify that the joint
system should be in the most entropic state, if you stay strictly
in the classical domain.

In standard statistical mechanics one would justify
maximizing the entropy of this joint system by considering all
possible fluctuations and arguing that, most of the time, the
system remains at or near the state maximizing the entropy.
However, your argument uses the area theorem, which requires that
the area never decrease!  Thus, if there is ever a fluctuation to
a state in which the area is {\it greater} than occurs in the most
entropic state, doesn't your argument forbid the system from ever
returning to the most entropic state? So then, why should the most
entropic state (of the surface and exterior together) characterize
this system?

\bigskip

{\bf Carlo:} Maybe I have over-emphasized the role of the classical
second law in determining the ensemble.  The question is whether in
the limit in which we disregard quantum effects, the area is
compatible with the Einstein dynamics, as a parameter characterizing
an equilibrium state.  The answer is yes in the sense that a change of
area is irreversible in classical GR. Hence an ensemble can be in
equilibrium only if it is at fixed area.  Maybe a classical
equilibrium ensemble is necessarily trivial, namely formed by a single
state, as your intuition suggests.  I am not sure, but even if you are
right, this is not the point here.  The ensemble I am interested in,
is not in the classical theory.

I know I am zigzagging between the classical theory and bits of
information about quantum effects.  Maybe this zigzagging is nonsense.
But this is the best I understand of black hole statistical physics.
I think and hope that a fully quantum treatment could fully justify
the legitimacy of this semiclassical thinking, but I cannot prove
this.

\bigskip

{\bf Ted:} OK, well let's put this question aside, and continue
the discussion of your derivation of the first law that you
sketched previously. It seems crucial there that (i) an {\it
energy} $dE$ is being transferred from the exterior to the surface
system, and (ii) the {\it energy} of the exterior decreases by the
same amount that the {\it energy} of the surface system increases.
It is only this way that the equality
$\frac{dS_{e}(E_{e})}{dE_{e}} = \frac{dS_{s}(E_{s})}{
dE_{s}}\equiv \frac{1}{T}$ follows, which allowed you to identify
the macroscopic parameter $T$ and its role in the first law for
the surface system. But $E_s$ is not energy, and $E_e + E_s$ is
not conserved. So, while I appreciate your later {\it description}
of the first law in terms of the change of ensemble over which the
horizon fluctuates, I don't see how you are offering a {\it
microscopic derivation} of this law as Don requested.

\bigskip

{\bf Carlo:} What I need to do, as I argued before, is to replace
$E_{s}$, which is what I called the ``energy-like" macroscopic
parameter of the surface, with the area $A$.  Then I call
$S_{s}(A)$ the logarithm of the number of states with area $A$.
The input I need from the Einstein equation is that I can increase
$A$ by an amount $dA$ by sending an amount $dE$ of external energy
over the horizon.  I know this happens for one particular
microstate: Schwarzschild. Therefore I know I can go from one
equilibrium state to the next by using this amount of external
energy.  The relation between $dA$ and and $dE$ in the case of
Schwarzschild is obtained from the relation between the energy
$E$, or mass $M=E$, of the quasi-isolated hole and its area
$A=16\pi G^2 E^2$.  Hence $dA=32\pi G^2 E dE$.  Thus, we can say
that an equilibrium statistical state with external energy $E-dE$
corresponds to the microcanonical surface ensemble characterized
by the macroscopical parameter $A+dA=A+32\pi M G^2 dE$.  Using the
line of argument above (which, as you said, is an adaptation of a
standard microcanonical argument), we have then that in order for
the system to be in a maximally probable state, the energy
transfer should not increase the total number of states, hence
there must exist a quantity $T$ such that
$\frac{dS_{e}(E_{e})}{dE_{e}} = 32\pi M G^2
\frac{dS_{s}(A)}{dA}\equiv \frac{1}{T}$.  Hence we have $dS=
dA/(32\pi M G^2 T)$.  Isn't this the first law?

Of course, in order to have the relation between this $T$ and $A$,
we have to know something else.  Either we use the indirect result
by Hawking that $T=\hbar/8\pi G M$ or, if you have some confidence
(as I do) in a quantum theory of gravity where you can actually
count the number of surface states with given area, you can
directly obtain this relation from the knowledge of the function
$S(A)$.

There are various weaknesses\ldots but to me this sounds like a
credible story for thinking about the black hole entropy in terms
of surface states in a fully general covariant setting, and
without being vulnerable to Don's tough criticism\ldots

\bigskip

{\bf Don:} Well, Carlo, I still have several other questions
\ldots For the first one, let's return to your ``replacement'' of
$E_s$ by the area.  Classically, it is true that if I limit myself
to moving from one Schwarzschild black hole to another, then
indeed $dE$ and $dA$ are related as you state above.  But more
generally this relation is modified.  For example, for Kerr black
holes there is also a term $\Omega dJ$.  And what if we consider a
static black hole which is distorted by some external matter, such
as a steel cage built around the black hole?  Then surely the
relation depends on many other parameters that describe this
cage...

\bigskip

{\bf Carlo:} The essential point of my discussion is the
distinction between the macrostate and the microstates.  So, you
are actually posing two different questions here.  The first
question refers to the macrostate.  Namely, you are asking what
happens if I am dealing for instance with a hole that has a
\emph{macroscopic} angular momentum. Or is \emph{macroscopically}
distorted, as a hole in a cage.

\bigskip

{\bf Don:} Actually, I would like to ask this question at both
levels, but let us discuss the macroscopic level first.

\bigskip

{\bf Carlo:}  Perfect.  Well, as is always the case in
thermodynamics, if you describe a system characterized by more
macroscopic parameters, such a rotating or distorted balloon,
you'll have a different ensemble. So, in a rotating or distorted
black hole, the ensemble will be described by more parameters than
just the area.  I haven't worked out these cases.  There is recent
work in this direction.\footnote{Carlo refers to \cite{abhay}.}
Here, for simplicity, will you allow me to consider just spherical
equilibrium states for the rest of our discussion?  It is like
doing the statistical mechanics of the gas in a spherical balloon.

\bigskip

{\bf Don:} Fine, but now let us move on to the microscopic version
of this question \ldots Do you require all of your microstates to
be spherical?  Birkoff's theorem helps you a lot in this
case\ldots

\bigskip

{\bf Carlo:} ({\em Much agitating hands in Italian way}) Not at
all! Not at all!  In a spherical balloon the motions of the
molecules  are {\em not} spherical!  If you try to work out the
statistical mechanics of a spherical balloon of gas by restricting
to spherical motions of the molecules you get it all wrong, of
course.  The fact that the hole is \emph{macroscopically}
spherical does not imply that its fluctuations are spherical!

\bigskip

{\bf Don:} Very good!  So this takes us back to my question: for
the large majority of your microstates $dE$ and $dA$ are not
related as you state above.

\bigskip

{\bf Carlo:} You are right!  Here is a subtle point.  Suppose I
have a system described by an ensemble of area $A$ and I want to
move it to an ensemble of area $A+dA$.  How much heat do I need?
Well, I do know that there is at least one path from one
microstate in the first ensemble (Schwarzschild) to one microstate
in the second ensemble (Schwarzschild with larger area) that can
be taken with cost $dE=dA/32\pi M G^2$.  Hence I know that this
amount of heat is sufficient to go from one ensemble to the other.

\bigskip

{\bf Don:} OK, I am still confused.  We agree that the change $dA$
in the horizon area depends on how you remove an energy $dE$ from
the exterior \ldots So, it looks to me like adding a given amount
of ``heat'' to one of your ensembles of fixed area does {\it not}
lead to another ensemble of fixed area \ldots Isn't this a
problem?

\bigskip

{\bf Carlo:} I am not saying that each microstate
in the first ensemble becomes a microstate of the second ensemble
by means of the same external $dE$.  I am saying that there is one
microstate with this property.  Therefore I expect that there
exists an adiabatic process going from one ensemble to the other
with a cost $dE$ in external heat.

\bigskip

{\bf Don:} Hmmm \ldots  I must say that my own
expectations differ markedly \ldots but let's move on another of
my questions.

Why, in your picture, does one maximize the joint probability only
for the surface system and exterior, without regard for the
interior. Don't we in general need to include all systems which
exchange, in your language, ``energy-like quantities"?  At least
quantum mechanically via Hawking radiation, energy certainly flows
back and forth between the surface system and the interior, so
doesn't this indicate that we should include the interior of the
black hole as well?

\bigskip

{\bf Carlo:} The condition $dS(E_{1})/dE_{1} = dS(E_{2})/dE_{2} =
1/T$ between two systems at equilibrium holds just because of the
interactions between these two systems.

\bigskip

{\bf Don:} Well, there is still the important question of whether
$dE_1=dE_2$ \ldots But I see that you expect the ``quantum first
law of black hole mechanics'' to take care of this issue...  Of
course, I would like to see the details \ldots

\bigskip

{\bf Carlo:} Yes, I am using a classical result, while I am already
outside the classical picture\ldots This is again the zigzagging
between classical and quantum I just referred to\ldots The entire
picture makes sense only under the unproven hypothesis that this
zigzagging is legitimate\ldots

\bigskip

{\bf Don:} OK, I think I now have a much clearer
understanding of your picture.  However, I see that you have fallen
back on logic much like Ted's, where you assume at least the classical
first law of black hole {\it mechanics}, which is what I had
implicitly hoped that you would derive.

\bigskip

{\bf Ted:} I object, your honor!  The logic is most certainly {\it
not} much like mine!  I {\it derived} the first law of black hole
mechanics, and the Einstein equation along with it!

\bigskip

{\bf Don:} Calm down now, Ted.  We've already agreed that you have
postulated a mechanism (whose existence you believe is associated
with the second law of thermodynamics, but whose microscopic
description you do not understand) to allow your surface system to
``keep" the entropy it picked up previously from a bit of energy
that was just passing through...  But I'll retract my reference to
your logic so we can focus on the issue at hand, which for now is
Carlo's use of the first law of black hole mechanics.

\bigskip

{\bf Carlo}: The story I am suggesting is precisely the
microscopic mechanism to allow the surface to ``keep" the entropy
it picks up when a bit of energy passes through: when the bit of
energy passes through, the surface keeps fluctuating
statistically, but over an ensemble of surfaces of larger area,
because this is the only possibility consistent with the Einstein
equation and the assumptions about macroscopic equilibrium.

But there is a major difference between what Ted does and what I
am doing.  Ted has this marvelous acrobatics that has allowed him
to actually {\em derive} the Einstein equation thermodynamically.
When he did this a few years ago, I remained astonished, and I
haven't yet digested the result.

I myself am far more conservative: I use the very standard logic
of statistical mechanics, where you assume mechanical laws, and
use statistical arguments to infer the thermodynamical ones.
Since gravity is involved, and its degrees of freedom are part of
the story, the relevant mechanical law is given by the Einstein
equation.

Therefore of course I use the laws of hole mechanics!  These laws
are just a consequence of the dynamics of the theory, of the
Einstein equation.  In fact, they are {\it the} property from
which the thermodynamical behavior follows.  They play the same
role as the microscopic energy conservation in the usual
nonrelativistic thermodynamic context.  I did not think that you
wanted us to derive {\em this} law!

\bigskip

{\bf Don:} Actually, I did\ldots You see, I was looking for
something more here...

As has often been noted, the laws of black hole mechanics strongly
resemble the laws of thermodynamics. And, one of the greatest
advances of physics was to understand that the laws of
thermodynamics follow (more or less) from elementary statistical
mechanics.  Thus, the laws of black hole mechanics are typically
taken to suggest underlying statistical properties of black
holes...  At least in my own view, this naturally suggests that
one should be able to use stat mech to derive the first law of
black hole mechanics {\it without} first assuming the Einstein
equation!  You may feel that this is too much to ask, either in
principle, or at least at our current state of knowledge, but it
is a goal to which I think we should aspire!

\bigskip

{\bf Carlo:} I think that you are mixing two different issues
here. One issue is whether a theorem that follows from the
Einstein equation can be derived from some statistical argument. I
have nothing to say about that.  Ted might see further than me
here, but at the present stage of knowledge, I prefer to take the
gravitational field as a field, whose mechanics is described, at
least within some approximation, by the Einstein equation.  The
other issue is whether or not entropy counts surface states and
how to derive $dS=dE/T$ from the surface state picture.  This
second issue is what I have discussed.  The two issues are quite
distinct.

\bigskip

{\bf Don:} Perhaps this suggests the crucial difference between
our points of view.  Of course the {\it classical} first law of
black hole mechanics follows directly from the classical Einstein
equation.  The same is true of the Hawking area theorem; i.e., of
the classical second law of black hole {\it mechanics}.  But we
know that the latter fails quantum mechanically, as the area will
decrease during Hawking radiation.  However, because black hole
entropy dominates over any other form of entropy as $\hbar
\rightarrow 0$ we can naturally {\it explain} the Hawking area
theorem as the classical limit of a second law of thermodynamics.
Such an explanation would work even if the quantum dynamical laws
are very different from the classical Einstein equations (perhaps
in analogy with the situation for hydrodynamics). To me, it seems
natural to seek a similar explanation for the {\it first} law of
black hole mechanics.  The {\it classical} first law of black hole
mechanics looks to me to be just the $\hbar \rightarrow 0$ limit
of the first law of thermodynamics.

\bigskip

{\bf Carlo:} Mmm\ldots I see nothing mysterious in the fact that
dynamical equations can be obtained as limits of quantum or
statistical equations.  This is just a minimal consistency
requirement.

Don, you are giving me your speculations on how you think Nature
works.  I listen with interest, but I see no relation with the picture
I have given.  I do not know if the gravitational field is a
fundamental field or not.  What I know is that there are some degrees
of freedom that in some approximation are well described by the
Einstein dynamics and, under the hypotheses of the picture I am
describing, that their statistical mechanics describes black hole
entropy.  The rest is your speculations, not mine!

\bigskip

{\bf Don:} But what is then your point of view?  You have only
mentioned the classical theory so far.  I get the feeling that you
expect the quantum dynamical laws to be much like their classical
counterparts...

\bigskip

{\bf Carlo:} I do.  In the sense in which a quantum harmonic
oscillator is ``much like" a classical oscillator.

\bigskip

{\bf Don:} \ldots and therefore that this ``quantum Einstein
equation" should lead to a ``quantum first law of black hole {\it
mechanics}" through an argument similar to that used to derive the
classical first law from the classical equations of motion\ldots

\bigskip

{\bf Carlo:} I am not sure, but I do not expect quantum
corrections to be much relevant in this \ldots

\bigskip

{\bf Don:} \ldots So, at the quantum level you would 1) find some
way to count states to make $A/4G$ into an entropy and then 2)
perform some quantum form of Hawking's calculation to show that a
quantum surface gravity $\kappa$ is associated with radiation at a
temperature $T = \hbar\kappa/2\pi$\ldots

\bigskip

{\bf Carlo:} As you know, states of a surface with given area are
counted in loop gravity.  The key point that makes the picture I am
describing meaningful is that the number of ``shapes" with a given
area is infinite in classical theory, but it turns out to be finite
and proportional to the (exponential of the) area in quantum gravity. 
But let's not discuss this here.  Calculations to derive Hawking's
radiation are in course, but I am not up to date.

\bigskip

{\bf Don:} \ldots this would then turn your quantum first law of
black hole mechanics into the (quantum) first law of
thermodynamics for black holes.  Is this indeed your opinion?  If
so, do you expect that the 2nd law can be derived the same way?

\bigskip

{\bf Carlo:} I prefer not to be drawn into speculations about the full
quantum statistical mechanics of gravity.  In order to understand
thermal quantum gravity we have to use a fully background independent
framework.  To formulate {\em classical} statistical mechanics in a
background independent language is hard enough.  Background
independent {\em quantum} statistical mechanics hasn't been much
explored yet, as far as I know.  I am convinced that this is where the
interesting problems are, and what we should do.  My attempt is a
first step in this direction, trying to develop steps of classical
background independent statistical reasoning\ldots For the full
quantum theory, I wait for some truly brilliant young kid to appear
and do it!  For the moment, we do not even know the basic grammar of
this.  We keep thinking in terms of stuff over a background: I am
convinced that this is wrong.

\bigskip

{\bf Don:} It may well be true twe need some conceptual shift away
from working around a background \ldots But, nevertheless, I feel that
you are avoiding the main question.  While I am not as sure as Ted
that, say, the gravitational field must in general be like a
hydrodynamic quantity, I do find compelling the idea that the
hydrodynamic analogy should apply {\it at least} to certain properties
of black holes.  \bigskip

{\bf Carlo:} Maybe ``analogies" can be found, but my expectation
is that it must be possible to understand black hole
thermodynamics entirely in terms of the quantum and statistical
behavior of the gravitational field, without any {\em need} of
resorting to some `hydrodynamical' approximation.  But if your
criticisms to the picture I have given is only that it does not
make use of these analogies, then I am quite happy!

\bigskip

{\bf Don:} Well, I trust you haven't forgotten the other issues that
Ted and I have raised \ldots

\bigskip

{\bf Carlo:} I haven't, Don!  Don't worry!

In fact, it has been a wonderful discussion and a lovely dinner \ldots
but now I need to move my legs.  How about moving to the lounge for an
after-dinner drink?

\bigskip
\begin{center}
------------------------------------------------------------
\end{center}
\bigskip

\begin{center}
{\bf  Scene III: States inside the lounge}
\end{center}

\bigskip

{\it The discussants have moved to a candle-lit lounge. Glasses of
the local digestif, g\'en\'epi, are poured.}

\bigskip
\begin{center}
------------------------------------------------------------
\end{center}
\bigskip

\bigskip
{\bf Don:} Well, it has been an intense day of discussion!  You
two are quickly wearing me out...  However, there is one more
point, Ted, on which I would like to press you.

Suppose we consider a setting where we maintain the black hole at
constant size (and prevent evaporation) via a steady influx of
radiation. Let's consider again the horizon-crossing surfaces in
my diagram\footnote{Recall that these are the hypersurfaces shown
on the right in Figure 1.} and note that the part of any such
surface that lies outside the black hole forms a Cauchy surface
for the exterior region.

In this setting an infinite number of Hawking quanta are
eventually emitted.  But all of these Hawking quanta can in some
sense be traced back to any horizon-crossing Cauchy surface, and
in particular to the part of this surface very near the horizon.
This is precisely your surface system. Now you claim that the
black hole entropy is really just the entanglement entropy of the
surface system, so that the surface system has a {\it finite}
entropy $S_{BH}= A/4\ell_p^2$, and thus a finite number of states.
How then can this finite number of states be consistent with the
radiation of an infinite amount of entropy via Hawking quanta?

\bigskip

{\bf Ted:} Yikes, you're right! But I propose an escape from this
conclusion: the standard field theory account of the Hawking
effect is unphysical on this point. If there really is a cutoff of
the entanglement entropy, and hence a cutoff of the density of
outgoing field modes at the horizon, then the outgoing modes
cannot arise from a trans-Planckian reservoir at the horizon, but
must rather come from elsewhere. I believe that the source is mode
conversion from ingoing to outgoing modes near the horizon, as
illustrated in various linear field theory models that have been
studied\footnote{Ted refers to models reviewed in  \cite{river}.},
including especially Hawking radiation on a falling lattice.

\bigskip

{\bf Don:} This sort of mode conversion violates Lorentz
invariance, and therefore invalidates one of the assumptions used
to deduce the thermal nature of the local Rindler horizon. So if
you really take this tack, aren't you now in trouble? Even if I
grant you the point we shelved earlier, how do you now argue that
the surface system is in equilibrium?

\bigskip

{\bf Ted:} Indeed the picture of vacuum conversion outside the horizon
is quite Lorentz violating, and yet in the free fall frame it looks
like the usual vacuum. This is why the studies of the Hawking effect
in this context have found that the usual Hawking effect is recovered.
Hence I think the thermality, while no longer exact to arbitrarily
high frequencies, is correct up to the cutoff.

\bigskip

{\bf Don:} But since your story is that the entropy is dominated by
the degrees of freedom at the cutoff, and it is just there that one
might expect thermality to break down, are you not in danger of
loosing the
canonical ensemble just where you most need it?

\bigskip

{\bf Ted:} Yes, I am in danger. I need another shot of
g\'en\'epi\ldots Ah, that's better. OK, the question is critical:
is the mode conversion vacuum outside the black hole horizon a
canonical ensemble all the way up to the cut-off? My guess is that
the answer is yes, but I can't assert that with total confidence
now. Within the linear field theory models it should be
straightforward to investigate the question. I'll do that. Suppose
it works out to be appropriately thermal. What will be your
response?

\bigskip

{\bf Don:} ({\it smiling})  Then I will certainly be impressed,
but I will still wait for you to fully derive the first law!

\bigskip

{\bf Ted:} So, Donny boy, after all this, we haven't convinced you
yet, eh?
Do you still think that black hole entropy has something---or
everything---to do
with the states {\it inside} the horizon?

\bigskip

{\bf Don:} You mean you're not totally devastated and ready to
convert to string theory? ({\it Don smiles mischievously.})

\bigskip

{\bf Ted:} No! In fact I'm more convinced than ever that  I'm right!

\bigskip

{\bf Carlo:}  I still see the surface picture as the most
convincing one, in spite of what I do not understand.  Especially
since I see no truly reasonable alterative.  Don,
show your cards: what is your view?

\bigskip

{\bf Don:} Well, I have to admit that I am attracted to your
approach. In particular, there is a result from the paper
with Minic and Ross that we've already mentioned which
has been on my mind of late.  There we found that two observers,
one who falls into a black hole and one who remains outside, do
not in general agree on the amount of entropy that a given object
carries with it into the black hole...  This does seem to be
suggestive that perhaps the entropy of the black hole itself is
somehow ``different'' as measured by observers inside and
outside...  but, unfortunately, I don't see precisely where to go
from here, and I keep getting hung up on the first law.  So, if
forced to choose, at the moment I think the more likely scenario
is one in which the Bekenstein-Hawking entropy counts the {\it
total} number of black hole microstates.  At least, this is what I
will take as my working hypothesis.

\bigskip

{\bf Ted:} Aha!  Then now it's our
turn to put you on the spot.
Can {\it you} give a statistical derivation
of the first law of black hole thermodynamics based on the idea
that black hole entropy counts the states on the {\it inside}?

\bigskip

{\bf Don:}  This, my friends, is child's play... The point is
that, if I assume that the {\it entire} black hole is described by
a standard quantum mechanical system, then I can use precisely the
usual stat mech argument to which Carlo referred earlier.  I place
my system in contact with some other stat mech system and let them
equilibrate.  This means that they randomly exchange energy back
and forth.

Now, the most likely macrostate for the joint system is the one
that maximizes the total number of microstates available. This
occurs when the transfer of a bit of energy $dE$ from one system
to the other leaves the total number of states unchanged.  Thus,
$dS/dE = 1/T$ must be the same for each system, and  the first law
($dE = T dS$) is the immediate result.  Here, I have merely given
a quick summary of the usual story from statistical mechanics...
Of course, I need all of the usual caveats about ergodicity, and
so forth.

However, before you both assault me physically, let me quickly
admit that there are {\it other} questions that are difficult in
this approach.  Notably, there is the point we have already
visited several times...  namely, that the classical physics of
the black hole interior does not in an obvious sense {\it look}
like it is approaching equilibrium.  Here I can only suggest that
this equilibration has something to do with the singularity.  To
make this fly, I probably need to mutter something about .... dare
I say it?...  non-local quantum gravity effects...

\bigskip

{\bf Ted and Carlo in unison:}  You can't be
serious!

\bigskip

{\bf Ted:} Why should causality fail so catastrophically inside a
black hole---even one that is a billion kilometers around the
horizon?!  Curvatures are small and semi-classical gravity should
apply deep inside.  Whatever happened to the equivalence principle??!

\bigskip

{\bf Carlo:} Yes !  Quantum gravity effects can very well play a role,
but at short scales around the horizon, not hugely non-local effects
connecting the singularity with the outside!

\bigskip

{\bf Don:} Hmmm...., I think I also need another shot of
g\'en\'epi....

I honestly don't know the answer to your question, but progress in
string theory over the past decade seems to indicate that this
picture is somehow consistent. Here to some extent I refer to the
stringy counting of BPS black hole entropy in terms of states on
D-branes, but what I really have in mind is Maldacena's AdS/CFT
conjecture.

\bigskip

{\bf Carlo:} Don, my dear friend, this is no fair!  I haven't mumbled
stuff like ``...but progress in loop quantum gravity over the past
decade seems to indicate that my picture is consistent".  This an
entirely different story!  If you have a credible story, even
incomplete, spit it out!  Otherwise, let's stay out from the fog and
from belief !

\bigskip

{\bf Ted:} Uh, oh.  Here we go again.  Don, you know my
arguments\footnote{Ted refers to his comments in \cite{TedAdS}.}
that AdS/CFT does not directly give us an answer to questions
about black hole entropy \ldots

\bigskip

{\bf Don:}  Yes, yes, but all I want to take from AdS/CFT is that
string theory in AdS space {\it does} seem to have something
fundamentally non-local about it, and that in some sense this
non-locality reaches out on a large scale all of the way to the
boundary of AdS space!  Yet, the theory appears to be consistent
and, at least in a classical limit, to provide local, causal, bulk
dynamics.  Thus, it seems plausible that a similar (though equally
unknown) mechanism could be at work in black holes.

I suppose the real point is that I take AdS/CFT as an indication
that spacetime itself will not be well-defined at a fundamental
level. As a result, I expect that causality will also fail to be
precisely well-defined, and thus that some apparently acausal
behavior may be allowed.  Actually, this seems like a natural
consequence of {\it any} theory of quantum gravity (like, say, one
of the standard versions of loop quantum gravity) which does not
explicitly build in a causal structure.

\bigskip

{\bf Carlo:} That spacetime itself is not well-defined at a
fundamental level is the working hypothesis of half a century of
research in nonperturbative quantum gravity, Don.  You know this
very well, of course!  Loop quantum gravity is just the attempt to
make this old idea sufficiently precise to compute with it.  Our
issue here is whether---if you have a macroscopic black
hole---whether acausal behavior can happen macroscopically.

\bigskip

{\bf Ted:} Yes, Don, while we don't expect classical causality
{\it per se} in quantum gravity, that does not give us license to
commit rampant violations of causality in an otherwise classical
setting.

I think there are two levels on which your
scenario requires causality violations. One relates to
correlations in Hawking radiation. If, as you believe, the entropy
{\it inside} a black hole is $A/4$, then a black hole maintained
at constant mass by an influx of energy in a pure state must emit
Hawking radiation in a pure state rather than the mixed state
predicted by semiclassical analysis. Since this semiclassical
prediction is based on analysis at scales very large compared to
the Planck length I don't think that quantum gravity effects can
invalidate it.

\bigskip

{\bf Don:} Well, while I have no idea how the details would work,
this seems quite plausible to me. After all, as has often been
emphasized, one only needs correlations between any two outgoing
Hawking particles that are exponentially small in the size of the
black hole \ldots  I'm referring to the fact that the non-zero
entries in a thermal density matrix are of order $\exp(-S)$, so
modifications of order $\exp(-S)$ could suffice to render the
density matrix pure \ldots I don't see how one can rule this out
without a complete theory of quantum gravity.

\bigskip

{\bf Ted:}  That seems quite {\it im}plausible to me. The
entanglement of each Hawking particle with a partner inside the
horizon is inferred from standard QFT vacuum structure at any
length scale much smaller than the black hole size, so is immune
to quantum gravity effects. Each such pair contributes of order
unit entropy, and I don't see how quantum gravity effects could
imprint the necessary correlations between the $M^2$ quanta that
are born one by one over a time of order $M^3$ and propagate
separately, far from the black hole out to the hinterlands. I'd be
very interested to see you fill in some details, even for a model.

\bigskip

{\bf Don:}  Yes, indeed, this would be a good
project \ldots

\bigskip

{\bf Ted:} However, there is another level on which your scenario
requires {\it far more} divine intervention. To support your view
that equilibrium is maintained inside a black hole, you require
not just a subtle quantum gravity effect, but rather total
breakdown of the causal {\it and} dynamical structure of
spacetime. You must replace the violent collapse to a spacelike
singularity---which is spread over eons of spacelike proper
distance---by some {\it equilibrium} dynamics!

\bigskip

{\bf Don:}  Sounds like a tall order, doesn't it? Well, perhaps by
fleshing out the ``surface system" point of view the two of you
can save me from such heresy... ({\it Smiles})

Of course, you have stated the main problem with this point of
view.  I really haven't thought enough about this point, but I
find myself wondering if it could be resolved by some appropriate
``coarse graining'' \ldots  Perhaps the non-equilibrium dynamics
inside the black hole is just a distraction, while ``most'' of the
black hole degrees of freedom equilibrate?  Susskind's ideas on
horizon complementarity\footnote{Don refers to e.g. \cite{LS}.}
would fall into this category, though I might hope for something
a bit more conventional \ldots   Hmm... It seems like I
would need to identify these degrees of freedom to pursue this
line of investigation.  Unfortunately, I have no concrete
suggestions at this time.

\bigskip

{\bf Ted:}  ({\it with a big grin})
May I suggest, perhaps, the degrees of freedom
just outside the horizon?

\bigskip

{\bf Don:} Your {\it suggestion} is duly noted \ldots as I said
earlier, I would be only to happy to see you fill in the relevant
details \ldots ({\it Smiles.})

My goodness, we've covered a lot of ground today\ldots  
I wonder if we should try to 
sum up where our discussion has left us?
It seems to me that we all have a lot of
work to do to before our pictures of black holes reach the level
that we would all truly like to see...

If I've got this right, then for Ted, its a matter of finding a
microscopic explanation---perhaps via the ``exterior second
law"---for why his surface system ``keeps" the entropy it gains
from each bit of energy that flows through.

\bigskip

{\bf Ted:} Ah, the second law, yes\ldots
maybe Rafael Sorkin's approach\footnote{Ted refers
here to \cite{RS1,RSGSL}.} can be made to work\ldots And there's
that other question: must I {\it really} invoke mode conversion to
keep the entanglement entropy finite, and if so can the canonical
ensemble for the surface system still be justified??

\bigskip

{\bf Don:} For Carlo, it remains to connect his mixed classical
and semi-classical statistical mechanics picture of an ensemble
with fixed area to a more complete quantum treatment.

\bigskip

{\bf Carlo:} I am happy with all this.  There is still confusion
under the sky: the situation is excellent.  We are still far from
the end of physics.

\bigskip  {\bf Don:} Agreed!
And finally, since I advocated
treating the Bekenstein-Hawking entropy as counting the number of
states in the interior of the black hole, I need some quantum
mechanism, perhaps violating our classical notions of locality and
causality, to explain how the black interior can be in equilibrium
and how the Hawking radiation can be in a pure state...

\bigskip

Ah, it's  good to see that we can have a
rational discussion about such things\ldots though I am sure that
the g\'en\'epi was essential\ldots ({\it Don smiles, Ted and Carlo
laugh.}) So, my friends, shall we call it a night? I am sure we
all look forward to our next discussion.... another place, another
time....

\bigskip

{\it  And so, our tired and by now slightly inebriated discussants
stumble out of the lounge and through the narrow medieval streets
of the village under the twinkling stars of the Provencal summer
night sky.  Each, of course,
thinking that they have had the best of the discussion....  but
each also having felt more sharply the shortcomings of their own
point of view.}

\begin{center}
------------------------------------------------------------
\end{center}

\section*{Acknowledgements}
We thank the organizers of the Peyresq 9 conference, 
which provided a stimulating atmosphere for our discussions. C.R. and
D.M. would also like to thank Larry Ford, with whom they discussed
related issues while departing from the conference.
 T.~J. was supported in part by
NSF grant PHY-0300710 and by the CNRS at the Institut
d'Astrophysique de Paris. D.~M. was supported in part by NSF grant
PHY03-54978 and by funds from the University of
California.

\end{document}